# Flexible One-Dimensional Metal-Insulator-Graphene Diode

Zhenxing Wang[1,*], Burkay Uzlu[1], Mehrdad Shaygan[1], Martin Otto[1], Mário Ribeiro[2], Enrique González Marín[3], Giuseppe Iannaccone[3], Gianluca Fiori[3], Mohamed Saeed Elsayed[4], Renato Negra[4], Daniel Neumaier[1]

[1]Advanced Microelectronic Center Aachen (AMICA), AMO GmbH, Otto-Blumenthal-Str. 25, 52074 Aachen, Germany

[2]CIC nanoGUNE, Tolosa Hiribidea 76, 20018 Donostia-San Sebastian, Spain

[3]Dipartimento di Ingegneria dell'Informazione, Università di Pisa, Via Caruso 16, Pisa 56122, Italy

[4]Chair of High Frequency Electronics, RWTH Aachen University, Templergraben 55, 52056 Aachen, Germany

[*]Author to whom correspondence should be addressed. Email: wang@amo.de

**Abstract**

In this work, a novel one-dimensional geometry for metal-insulator-graphene (1D-MIG) diode with low capacitance is demonstrated. The junction of the 1D-MIG diode is formed at the 1D edge of $Al_2O_3$-encapsulated graphene with $TiO_2$ that acts as barrier material. The diodes demonstrate ultra-high current density since the transport in the graphene and through the barrier is in plane. The geometry delivers very low capacitive coupling between the cathode and anode of the diode, which shows frequency response up to 100 GHz and ensures potential high frequency performance up to 2.4 THz. The 1D-MIG diodes are demonstrated to function uniformly and stable under bending conditions down to 6.4 mm bending radius on flexible substrate.

**Introduction**

Diodes are key components in radio frequency (RF) communication systems, where they are used for several RF applications like signal rectification, power detection, and energy harvesting, just to cite few. [1-5] Thin-film technology based diodes such as metal-insulator-metal (MIM) diodes or 2-dimensional (2D) material based metal-insulator-graphene (MIG) diodes are attracting increasing research interests over the past years, due to their high performance and thin-film fabrication process on rigid or flexible substrate. [6-16] Moreover, MIG diodes are also used as building blocks to realize different circuits, such as mixers, power detectors, RF receivers etc. [17-21] In MIM or MIG diodes, the junction is defined by an insulating barrier layer either between two metal layers with different work functions (in the case of MIM diodes) or between one metal layer and one graphene layer (in the case of MIG diodes). The barrier layer and work-function difference dominates the rectification behavior of the MIM or MIG diodes. However, in the conventional geometry the two metal layers form a parallel plate capacitor with a thin barrier in-between, which leads to a huge intrinsic junction capacitance. This results in large RC time constant, limiting the maximum operation frequency of a communication system, which is not preferable for future high-speed data communication.

Here we propose a radically new MIG diode geometry, which significantly reduces the junction capacitance, a so-called one-dimensional metal-insulator-graphene (1D-MIG) diode geometry. In this geometry, the graphene is encapsulated in a oxide layer and exposed only at the 1D edge, as shown in Figure 1a. On one of the two edges the graphene is contacted with an ohmic metal contact serving as cathode of the diode. [22] On the other edge of the graphene, the 1D junction is formed between the exposed graphene 1D edge and the metal electrode (the anode) separated by a thin insulating barrier. The carrier transport through the barrier layer takes place only at the 1D edge of graphene, which minimizes the capacitive coupling of the metal and the graphene. In addition, the carrier transport in the graphene layer and through the barrier layer happens in plane with graphene sheet, so the current density of the diode is expected to be high, which also ensures the performance at high frequency range.

**Experimental Section**

For the fabrication of 1D-MIG diodes commercial available chemical vapor deposition grown graphene and photolithography technology was used on rigid silicon substrate ($\rho$ > 5 k$\Omega$·cm with 1 μm $SiO_2$) and on flexible foil. Atomic layer deposition (ALD) is utilized to deposit the 40 nm $Al_2O_3$ encapsulation layer on top of the graphene. Reactive ionic etching (RIE) is used to pattern the $Al_2O_3$/graphene stack, and the graphene is contacted ohmically with metal Ni at cathode side. A $TiO_2$ layer of 6 nm from ALD is deposited to form the barrier layer for the diode on the anode side, and Ti is used as the anode metal. [15] 100 nm Al was deposited to form contact pads for probing. The optical microscope image of the device in Figure 1b shows the dimension of the diode. The graphene width is 70 μm, and the graphene length between anode and cathode is 12 μm. Detailed description and schematics (Figure S1) regarding fabrication could be found in the supporting information.

**Results and Discussion**

In order to gain insights on the electrostatic playing in the 1D-MIG diode, we have numerically solved the 2D Poisson equation in the y-z plane of the structure shown in Figure 1a, while assuming it invariant along the x direction (the numerical implementation is described in details in the supporting information). In particular, we focus on the region close to the 1D junction and we assume that the graphene-cathode junction is an Ohmic contact, as supported by experimental observations. [22] We use a linear dispersion relationship (Fermi velocity close to $10^6$ m/s) and Fermi-Dirac statistics to describe the carrier density in the graphene layer. As for the dielectrics, we have considered a dielectric constant κ = 3.9 for SiO$_2$, κ = 15 for TiO$_2$ and κ = 9 for Al$_2$O$_3$. The computed potential energy profile *vs.* the applied bias across a 1D cut along the graphene-TiO$_2$-Ti heterojunction is shown in Figure 1c, for different biases, i.e. -1 V, 0 V and 1 V (see supporting information for more details). The energy barrier is clearly asymmetric for different injection directions. For electrons injected from graphene, the barrier (i.e. the difference between the electrochemical potential of graphene, which has been set to 0 eV, and the top of the conduction band at the graphene/TiO$_2$ interface) varies with the bias voltage. [15] On the contrary, for electrons injected from the Ti cathode, the barrier height remains constant. For positive biases, the graphene/TiO$_2$ barrier is lower, ruling the electron transport; while for negative biases, the electron transport is controlled by the Ti/TiO$_2$ barrier. This asymmetry in the carrier injection has a direct translation into the current flowing through the 1D junction, which is consequently expected to show an asymmetric and non-linear behavior. The calculated junction capacitance ($C_j$) (calculated as indicated in the supporting information) is plotted in Figure 1d, with values from 0.155 to 0.268 fF/µm, depending on the applied bias, which correspond to 11 and 19 fF for a width of 70 µm, which was also used in the measurement later on. For comparison, in the conventional MIG diode, [15] where the graphene and the metal form a parallel plate capacitor, we measured a capacitance of 750 fF using the same device width of 70 µm and an overlap of 2 µm between metal and graphene. The expected RC bandwidth ($f_c$) of the 1D-MIG diode is

$$f_c = \frac{1}{2\pi \cdot R_a \cdot C_j} = \frac{1}{2\pi \cdot 6\,\Omega \cdot 11\,fF} = 2.4\,THz,$$

assuming that the access resistance ($R_a$) is 6 Ω, which is calculated by using an access length of 200 nm, a contact resistance of 200 Ω·µm [22,23] and a sheet resistance of 1 kΩ/□. This would place the 1D-MIG diode among the fastest diodes available.

To evaluate the 1D-MIG diode for the application in rectifiers or power detectors, the *I-V* characteristics was measured under ambient conditions (Figure 2a) and the typical figures of merit (FOMs) for diodes are extracted, in particular, the asymmetry of a diode defined as $f_{ASYM} = \left|\frac{J_F}{J_R}\right|$, where $J_F$ and $J_R$ are the current density for forward and reverse bias, respectively, the nonlinearity as $f_{NL} = \frac{dJ}{dV}/\frac{J}{V}$, where $J$ is the current density of the diode, and the responsivity as $f_{RES} = \frac{d^2J}{dV^2}/\frac{dJ}{dV}$. The *I-V* characteristic shown in Figure 2a exhibits a strong rectification behavior which is characterized quantitatively by $f_{ASYM}$. As shown in Figure 2b $f_{ASYM}$ increases for increasing biases, reaching a maximum value of 18 at bias of 0.9 V. The extracted $f_{NL}$ is a measure of the deviation of the diode performance from the linear behavior. As shown in Figure 2c, the nonlinearity increases with the bias voltage, reaching a maximum of 3 at 0.3 V. The maximum responsivity of 16 V$^{-1}$ is obtained at 0.1 V (Figure 2d), which quantitatively benchmarks the capability of a diode working as a power detector. All these FOMs demonstrate that the 1D-MIG diode performs excellently. Moreover, Figure 3 shows device to device variations of the 1D-MIG diode, with the histograms for the different FOMs. Additionally the histogram of the bias voltage, when the $f_{ASYM}$ reaches maximum value, is also shown in Figure S2. Since the process utilized is scalable, the yield of the device is higher than 90%. In total we have measured 50 devices and 49 of them perform properly. The histograms prove that this process is uniform in terms of FOMs. To calculate the current density, the cross-section is obtained from the product of the width of the 1D edge (70 µm) and the thickness of graphene (0.3 nm). The FOMs for diodes based on different thin-film technologies are benchmarked in Table 1. As can be seen from the table, the current density of 1D-MIG diode is extremely high compared to other technologies. The high current density does not only originate from the small thickness of graphene, but also because high absolute currents of the order of mA are measured. This is a significantly higher value compared to the conventional MIG diode with the same channel width and with 2 µm diode length, where 10 µA were measured at 1V forward bias. The high current level in the 1D diode is explained by the completely in plane current flow and the absence of a thin tunneling barrier as proposed for vertical devices. [24] The other FOMs of 1D-MIG are also comparable to other technologies in Table 1. Another significant advantage of such 1D-MIG diodes is that the graphene is encapsulated from the very beginning of the process flow, and therefore it is protected from any possible contamination from the environment and process, which makes the process more scalable, reliable and reproducible. Besides, the encapsulation could also ensure the long term stability of such devices. [25]

In order to understand the impact of the barrier on the diode characteristic, different materials and thicknesses for the barrier layer are investigated as control experiment. All the investigated devices are with the same dimension as the one shown in Figures 1. For a diode without any barrier material and for thin (2 nm) $TiO_2$ the devices behave completely linear (Figure 4a and 4b) with even higher current levels compared to the 1D-MIG diode shown in Figure 2a. In this case ohmic contacts are formed on both sides. If we use $Al_2O_3$ instead of $TiO_2$, either with 2 nm or 10 nm barrier, as shown in Figure 4c and 4d, rectification behaviors can be again observed but the current level is much lower (at least 3 orders of magnitude) compared to the diode with 6 nm $TiO_2$. This is because of the large band gap of $Al_2O_3$ significantly reducing thermal emission and leaving tunneling as the dominating transport mechanisms across the barrier.

The device is characterized with a network analyzer from 10 MHz to 70 GHz, in order to explore the RF performance and to extract the capacitance of the 1D-MID diode. One-port S-parameter measurement based on the layout shown in Figure 5a is carried out to measure the reflection coefficient $S_{11}$. There are two diode channels with 70 µm width each in parallel. The terminal G is grounded and bias was applied to the terminal S, by using a GSG probe. The measured $S_{11}$ over the whole frequency range is plotted in the Smith chart shown in Figure 5b for different bias voltages. The equivalent device model shown in Figure 5a was used to extract characteristic parameters including the junction capacitor ($C_j$), the junction resistor ($R_j$) and the access resistance ($R_a$). Based on this model, the parameters are extracted and plotted in Figure 5c. A series resistance $R_a$ is extracted as 16 Ω per 70 μm diode channel, which is constant over the whole frequency range. This series resistance is mainly related to the rather long access region, which could be significantly reduced. The measured capacitance of the diode after de-embedding is about 100 fF per 70 μm diode channel ($C_j/2$ in Figure 5c). Here for the de-embedding procedure we only utilized the big pads as an open de-embedding structure, so only the fringe capacitance between the pads as well as the parasitic capacitance via the substrate is eliminated. The device geometry capacitance introduced by the pads in the dashed box marked in Figure 5a is still adding a constant capacitance up to the intrinsic capacitance of the diode (which should be dominated by the bias dependent quantum capacitance). The rather slight dependency on the applied bias of the capacitance in Figure 5c also gives an indication on this. Therefore the measured capacitance here (100 fF) is higher than the simulated value (11 fF). However, the corresponding $RC$ bandwidth of the measured diode is about $(2\pi R_a C_j)^{-1} = (2\pi \cdot 16\ \Omega \cdot 100\ \text{fF})^{-1} = 100$ GHz. This is already outstanding and can be further increased if the access length of graphene is reduced, or the diode junction is geometrically optimized in terms of less parasitic capacitance in order to approach the theoretical value from the simulation.

1D-MIG diodes were also fabricated on the flexible polyimide (PI) substrate as shown in Figure 6a. The fabrication is the same as on rigid substrate and the yield is also higher than 90%. The *I-V* characteristic in Figure 6b shows good rectification behavior, with excellent reproducibility on 5 different devices, although it shows lower current compared to the devices on rigid substrate (Figure 2). The lower current could be explained by the RIE process on PI, which could generate some re-deposition, increasing the contact resistance at the edge of graphene. The bending test of the sample is also carried out. The strain applied to the substrate is defined by *T/2R*, where *T* is the thickness of the substrate, and *R* is the radius of curvature for bending. Since the thickness of the PI substrate is only 8 μm, it is challenging to apply relatively high strain. Nevertheless, different bending radii of 25.4 mm, 12.7 mm, 6.4 mm are applied. The diode performance (asymmetry $f_{ASYM}$ and ON current) with strain applied is shown in Figure 6c. It is clear that the device performance under strain is stable. Furthermore, we applied flex cycles for the substrate at bending radius of 6.4 mm. The diode performance (asymmetry $f_{ASYM}$ and ON current) shows no degradation up to flex cycles of 1000, as in Figure 6d. In this case the devices were measured in flat status after certain flex cycles. These tests also expand the application fields for such diodes to flexible electronics, e.g. wearable data communication systems.

**Conclusions**

A novel 1D-MIG diode with improved RF performance is proposed and demonstrated, exhibiting high current density and low junction capacitance ultimately enabling THz operation. A statistical investigation is carried out to prove the scalability of the process, which shows comparable performance as other thin-film technology based diodes, but much higher current density. The 1D-MIG diodes are also demonstrated to function uniformly and stably under stress on flexible substrate. The 1D-MIG diodes provide a solution for realizing high speed communication systems based on either rigid thin-film technology or flexible electronics.

**Acknowledgements**

This work was financially supported by the European Commission under the project Graphene Flagship (contract no. 785219), and the Marie Curie Actions (607904-13-SPINOGRAPH); and by the German Science Foundation (DFG) within the priority program

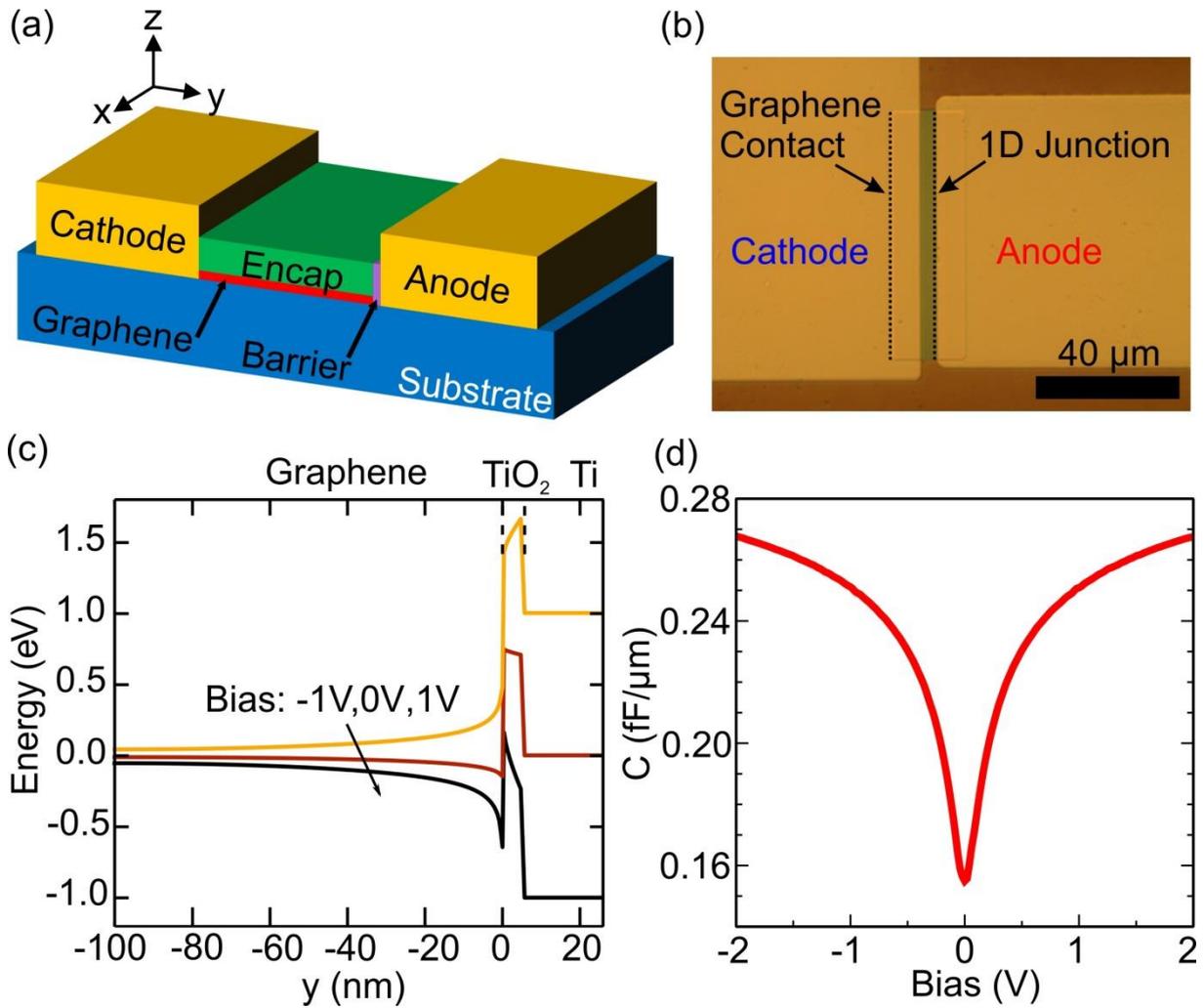

Figure 1. (a) The geometry illustration and (b) the optical micrograph of the 1D-MIG diode. The cathode is ohmically contacted with graphene from the 1D edge, and the anode forms a 1D MIG junction with graphene through a $TiO_2$ barrier. The graphene is encapsulated with $Al_2O_3$ layer. (c) Potential energy profile along the 1D junction for different biases. (d) Simulated bias dependent capacitance of the diode junction.

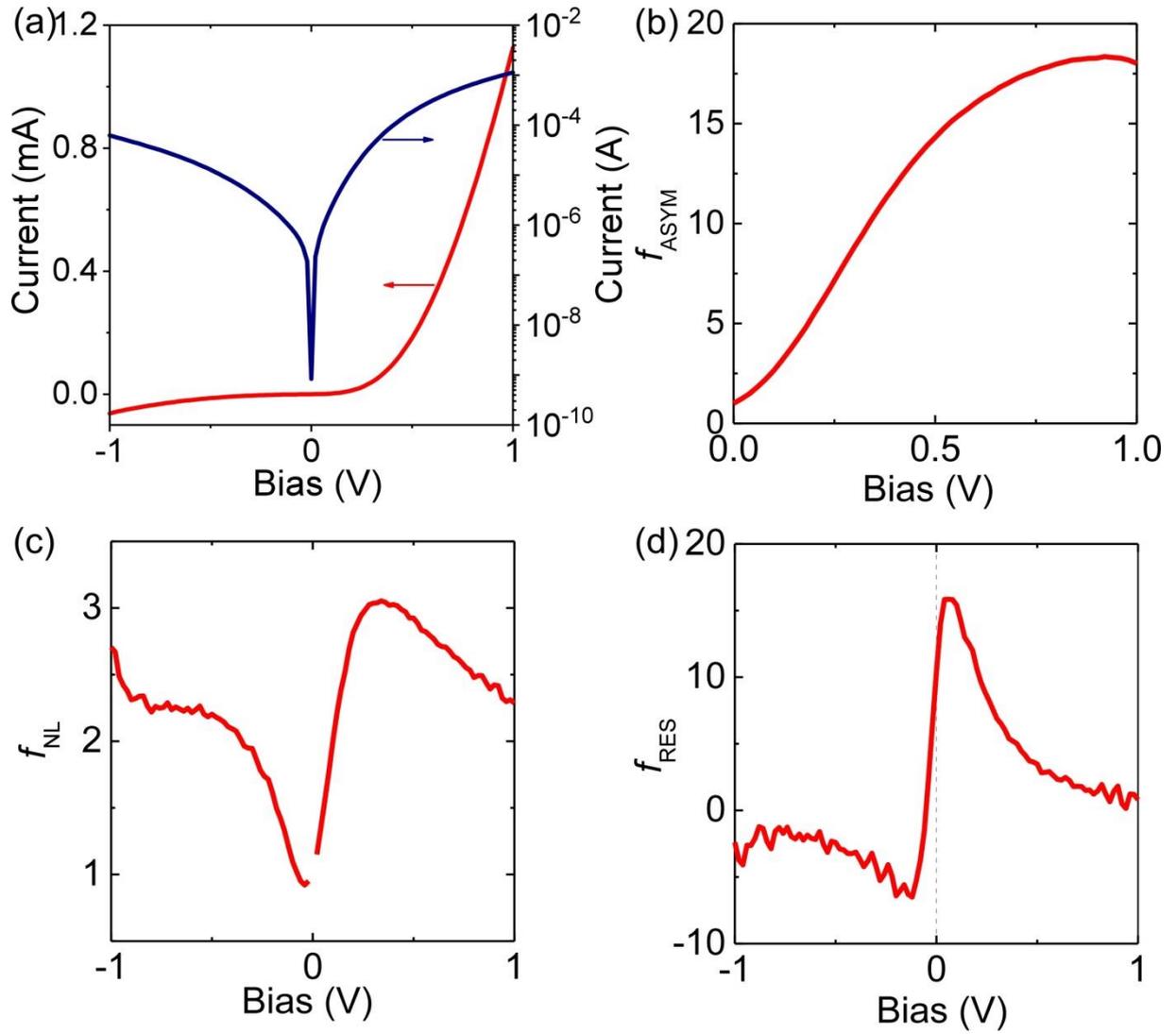

Figure 2. The DC characteristics of a typical 1D-MIG diode: (a) I-V characteristic, (b) asymmetry ($f_{ASYM}$), (c) nonlinearity ($f_{NL}$), and (d) responsivity ($f_{RES}$). These parameters are typically used as FOMs to characterize and benchmark a diode.

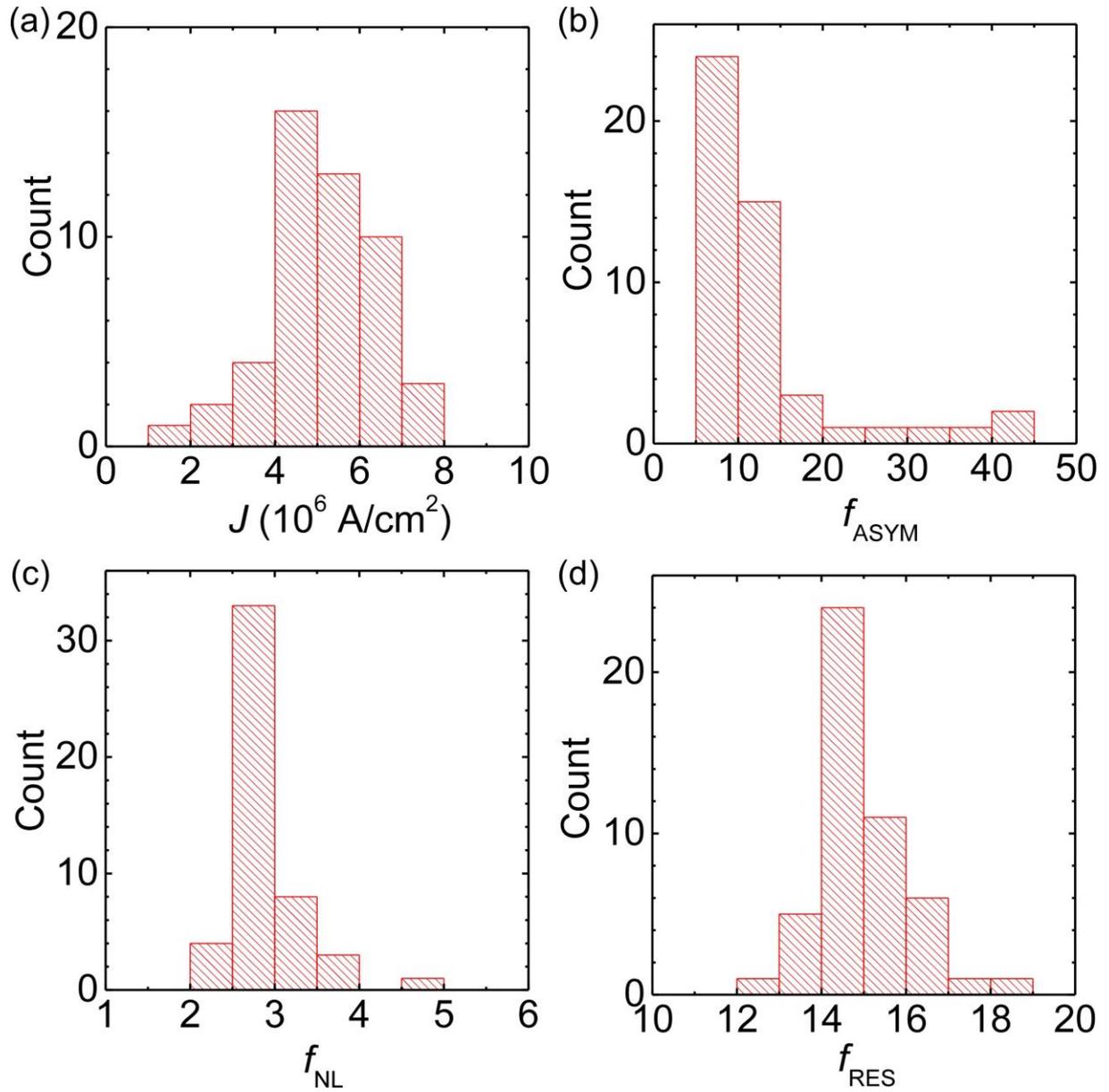

Figure 3. Histogram of the FOMs of diodes on a chip: (a) Current density ($J$), (b) asymmetry ($f_{ASYM}$), (c) nonlinearity ($f_{NL}$), and (d) responsivity ($f_{RES}$).

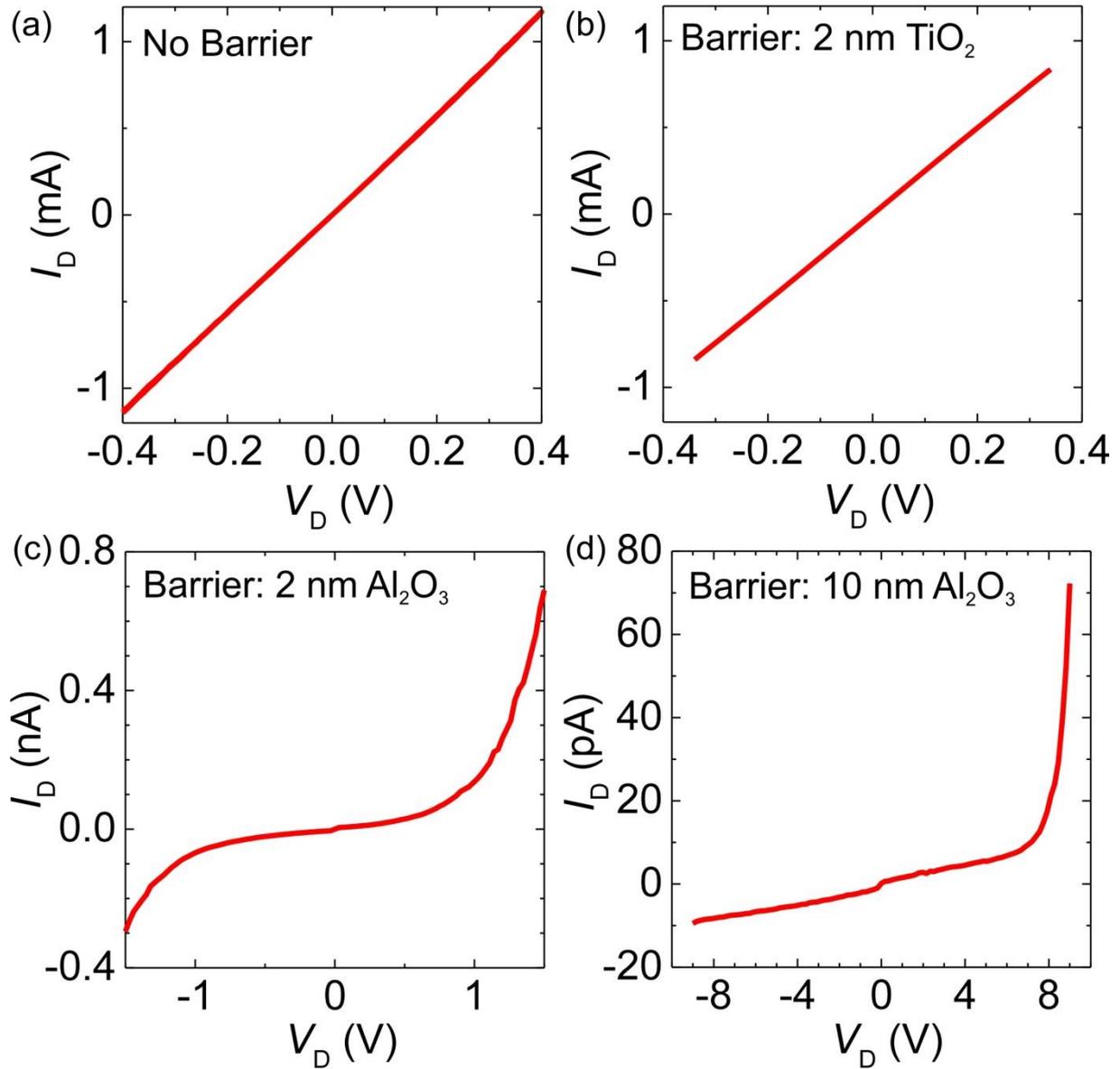

Figure 4. Control experiment to identify the thickness and the material of the barrier layer. Different configuration is tested with *I-V* characteristics: (a) no barrier applied, (b) 2 nm $TiO_2$ as barrier, (c) 2 nm $Al_2O_3$ as barrier, and (d) 10 nm $Al_2O_3$ as barrier. None of them shows comparable rectification performance in terms of current level.

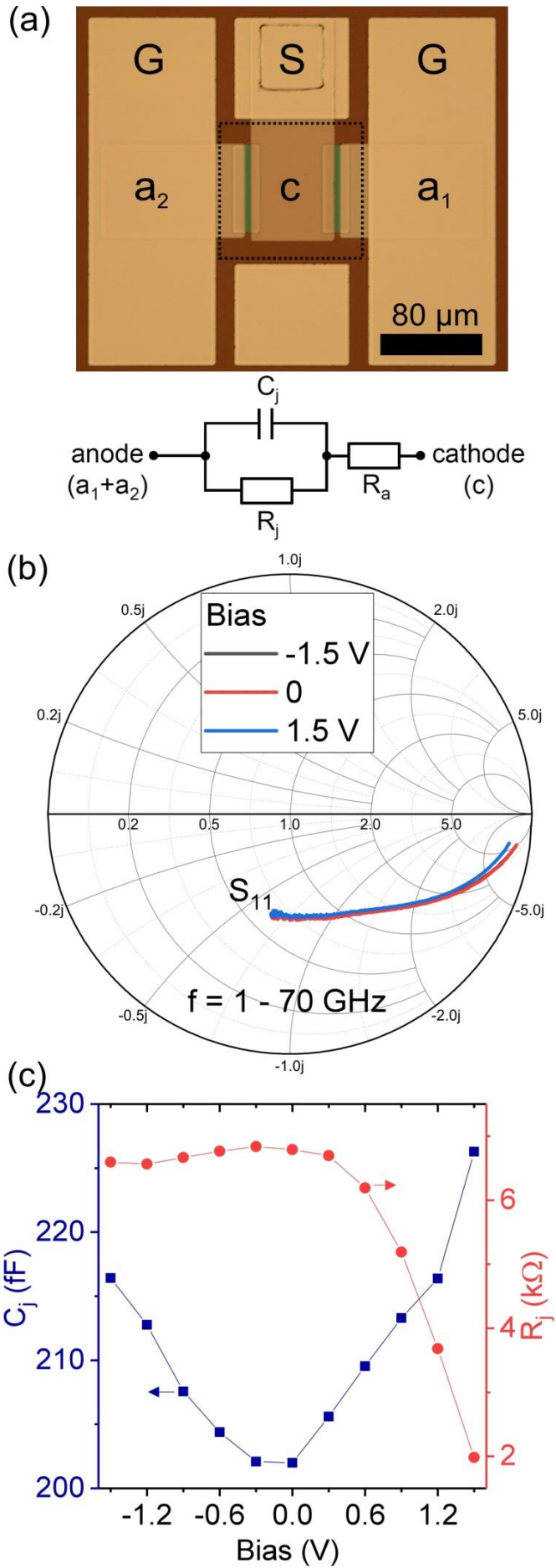

Figure 5. One-port S-parameter characterization of the 1D-MIG diode. (a) The GSG layout for the S-parameter measurement, as well as the used model for parameter extraction. The junction capacitance ($C_j$) and the junction resistance ($R_j$) are in parallel, with the access resistance ($R_a$) in serie. (b) Smith chart of $S_{11}$ with frequency sweep from 1 to 70 GHz, for different diode bias voltages. (c) The extracted value of junction resistance and junction capacitance for different diode bias voltages.

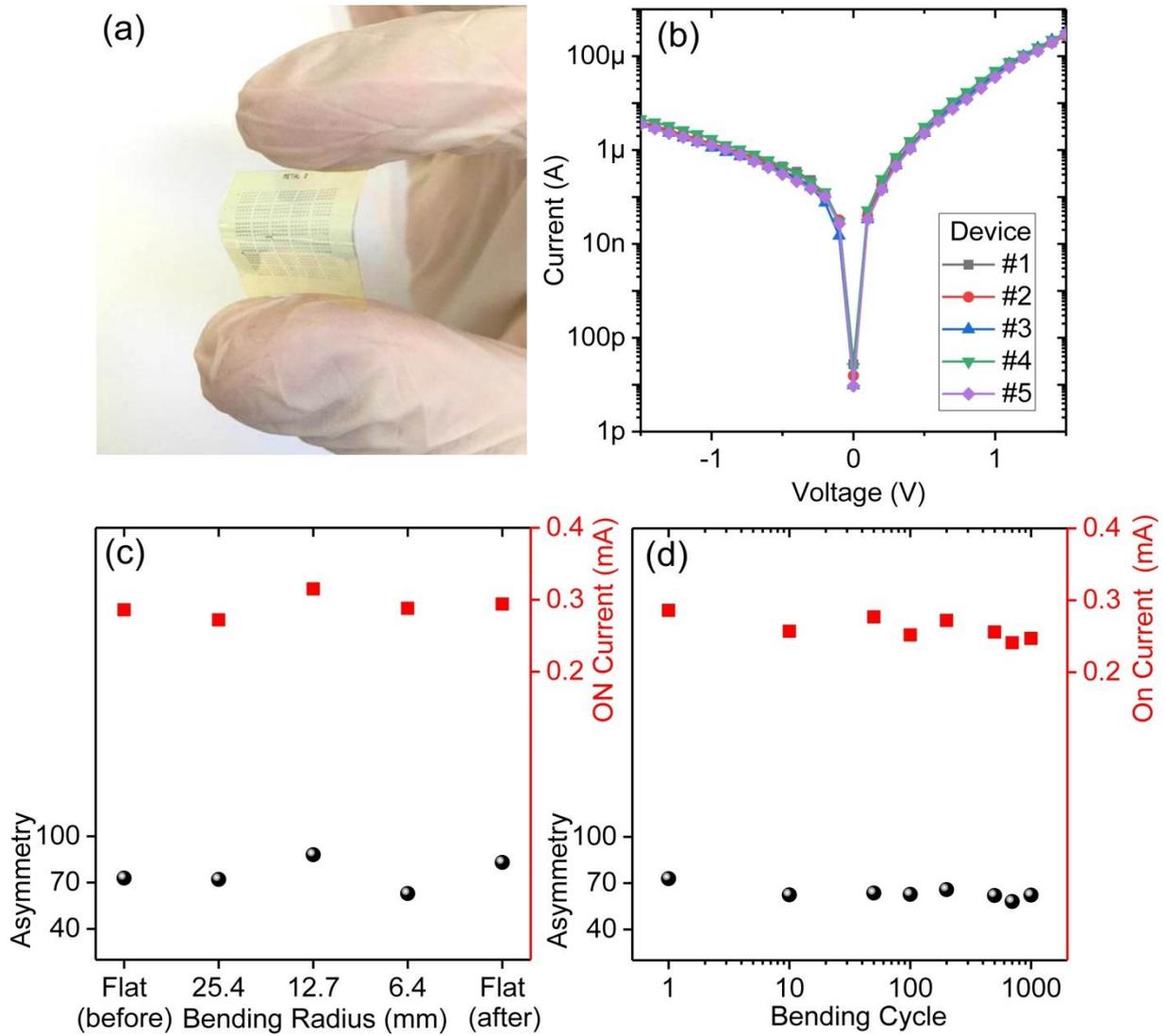

Figure 6. Flexible 1D-MIG diodes. (a) The photo shows the bendability of the chip with 1D-MIG diodes. (b) The I-V characteristics of the 1D-MIG diodes from five different devices. The uniformity of the device performance over the flexible chip is clearly demonstrated. (c) The stability of the 1D-MIG diodes while different strain applied. (d) Flex cycle test for the 1D-MIG diodes, the device is subjected to bending cycles with bending radius of 6.4 mm and is measured in flat status after the corresponding flex cycles.

Table 1. Performance comparison of 1D-MIG diodes with other thin-film technology based diodes. The current level is measured at bias voltage of 1 V, except for Reference 7, which is measured at bias voltage of 0.5 V. The current density is the value at bias voltage of 1 V.

| | $I$ (@1 V) | $J$ (A/cm$^2$) | $f_{ASYM}$ | $f_{NL}$ | $f_{RES}$ (V$^{-1}$) |
|---|---|---|---|---|---|
| Nb/Nb$_2$O$_5$(5 nm)/Pt [8] | 128 µA | 2 | 9.8 | 8.2 | 16.9 |
| Nb/Nb$_2$O$_5$(15 nm)/Pt [7] | 175 nA (@0.5 V) | - | 1500 | 4 | 20 |
| Ti/TiO$_2$/bilayer graphene [14] | 10 nA | 0.1 | 9000 | 8 | 10 |
| Ti/TiO$_2$/graphene (average) [15] | 5.3 µA | 3.8 | 320 | 12 | 24 |
| Ti/TiO$_2$/graphene (max.) [15] | 10.5 µA | 7.5 | 520 | 15 | 26 |
| 1D-MIG (average) [this work] | 1.1 mA | 5.1×10$^6$ | 14.1 | 2.8 | 15.0 |
| 1D-MIG (max.) [this work] | 1.6 mA | 7.5×10$^6$ | 65.7 | 4.5 | 18.5 |